\documentclass[twocolumn,english,pra]{revtex4}
\usepackage{ae,aecompl}
\usepackage[T1]{fontenc}
\usepackage[latin9]{inputenc}
\usepackage{color}
\usepackage{babel}
\usepackage{amsmath}
\usepackage{amssymb}
\usepackage{graphicx}
\usepackage{esint}
\usepackage[unicode=true,pdfusetitle,
 bookmarks=true,bookmarksnumbered=false,bookmarksopen=false,
 breaklinks=false,pdfborder={0 0 1},backref=section,colorlinks=true,pdfpagemode=None]
 {hyperref}
\hypersetup{
 linkcolor=blue,citecolor=black,urlcolor=blue,plainpages=false,pdfpagelabels,hypertexnames=false,pdfstartview=FitH}

\makeatletter
\@ifundefined{textcolor}{}
{%
 \definecolor{BLACK}{gray}{0}
 \definecolor{WHITE}{gray}{1}
 \definecolor{RED}{rgb}{1,0,0}
 \definecolor{GREEN}{rgb}{0,1,0}
 \definecolor{BLUE}{rgb}{0,0,1}
 \definecolor{CYAN}{cmyk}{1,0,0,0}
 \definecolor{MAGENTA}{cmyk}{0,1,0,0}
 \definecolor{YELLOW}{cmyk}{0,0,1,0}
}

\usepackage{aecompl}
\usepackage{color}\usepackage{babel}
\usepackage{ulem}

\usepackage{babel}
\usepackage{babel}

\makeatother

\begin{document}

\title{Coherent and Incoherent Multiple Scattering}

\author{Julien~\surname{Chabé}$^{1,2}$, Mohamed-Taha Rouabah$^{1,3}$,
Louis~\surname{Bellando}$^{1}$, Tom~\surname{Bienaimé}$^{1,4}$,
Nicola~\surname{Piovella}$^{5}$, Romain~\surname{Bachelard}$^{6}$,
and Robin~\surname{Kaiser}$^{1}$}

\affiliation{$^{1}$Universit{é} de Nice Sophia Antipolis, CNRS, Institut Non-Lin{é}aire
de Nice, UMR 7335, F-06560 Valbonne, France\\
 $^{2}$Laboratoire Lagrange, UMR 7293, Universit{é} de Nice Sophia-Antipolis,
CNRS, Observatoire de la Côte d\textquoteright{}Azur, Parc Valrose,
Bât. H. Fizeau, F-06108 Nice, France\\
 $^{3}$Laboratoire de Physique Math{é}matique et Physique Subatomique, Université Constantine 1, Route Ain El Bey, ALG-25017 Constantine, Algeria\\
 $^{4}$Laboratoire Kastler Brossel, CNRS, ENS, UPMC, 24 rue Lhomond,
F-75005 Paris, France\\
 $^{5}$Dipartimento di Fisica, Università degli Studi di Milano,
Via Celoria 16, I-20133 Milano, Italy\\
 $^{6}$Instituto de F\'{i}sica de São Carlos, Universidade de São
Paulo, B-13560-970 São Carlos, SP, Brazil}
\begin{abstract}
We compare two different models of transport of light in a disordered
system with a spherical Gaussian distribution of scatterers. A coupled
dipole model, keeping into account all interference effects, is compared
to an incoherent model, using a random walk of particles. Besides
the well known coherent backscattering effect and a well pronounced
forward lobe, the incoherent model reproduces extremely well all scattering
features. In an experiment with cold atoms, we use the momentum recoil
imparted on the center of mass of the sample as a partial probe of
the light scattering properties. We find that the force acting on
the center of mass of the atoms is not well suited to exhibit the
coherence effects in light propagation under multiple scattering conditions.
\end{abstract}


\maketitle

\section{Introduction}

Coherence effects in transport of wave in disordered systems are at
the heart of many phenomena in various areas of research. In this
work, we focus on light propagation, even though many if not most
features could be extended to different types of waves, be it acoustic
waves, plasmons, heat, antennas or matter waves such as electrons
or ultracold atoms. In mesoscopic physics, coherences are fundamental
for weak \citep{PhysRevLett.55.2692,PhysRevLett.55.2696} and strong
localization of light \cite{WiesmaLocLight97,Maret1999,Lagendijk1999,PhysicsToday,Maret2013a,scheffoldinelastic2013,maret2013inelastic,CFS_BvT,CFS_Miniat}.
Coherences are also at work in the universal conductance fluctuations
\citep{PhysRevLett.81.5800}, modifications of the local density of
states \cite{Carminati} or extraordinary optical transmission \cite{Lalanne}.
Cooperative emission of light as discussed by Dicke in the 50s \cite{Dicke1954}
and the response of a cloud of cold atoms excited by an external laser
\cite{Courteille2010} are also based on coherence effects, such as
quantum memories using electromagnetic induced transparency in three-level
systems \cite{PhysRevLett.83.1319}. The recent development of ultrastable
atomic clocks also relies on optical transitions in presence of many
atoms \cite{Ye_Science2014} where the impact of residual multiple
scattering deserves particular attention.

Whereas a rigorous investigation of light propagation in the presence
of many scatterers requires taking into account the effects of interferences,
in most situations interference effects can be neglected and a radiative
transfer equation is thus often used in optics, allowing a practical
approach to scattering of light in complex media. In this work, we
compare an approximate model, based on such an incoherent random walk
of photons, to a more rigorous approach, based on a microscopic coupled
dipole model. Cold atoms provide an excellent medium to study these
fundamental effects. This ensemble of resonant point scatterers are
free of defects and absorption and the coupled dipole model is expected
to provide an excellent description of the scattering properties of
this sample (despite some limitations, which will be discussed at
the end of the work). We stress that despite the apparent simple situation,
no analytical result is available for a disordered system of $N$
coupled dipoles. This problem has the full complexity of a true many-body
problem \cite{RevModPhys.70.447} and one thus needs to resort to
numerical or experimental answers to this question. Indeed, if the
amplitudes of $N$ coupled dipoles are to be found, this amounts to
solving $N$ coupled equations, even though this can be considered
as a linear optics problem described by the propagation of a low intensity
or single photon field.

This paper is constructed as follows. First we present the results
of numerical simulations using both a Random Walk of photons (RW)
and a Coupled Dipole approach (CD), where the many-body problem is
solved by tracing over the photon degrees of freedom. Then we compare
the numerical results from both the RW and CD models to experimental
data obtained by monitoring the radiation pressure force on the center
of mass of the atomic cloud as a probe of the emission diagram.

\section{Random Walk Model}

\begin{figure}
\centering{}\includegraphics[width=1\columnwidth]{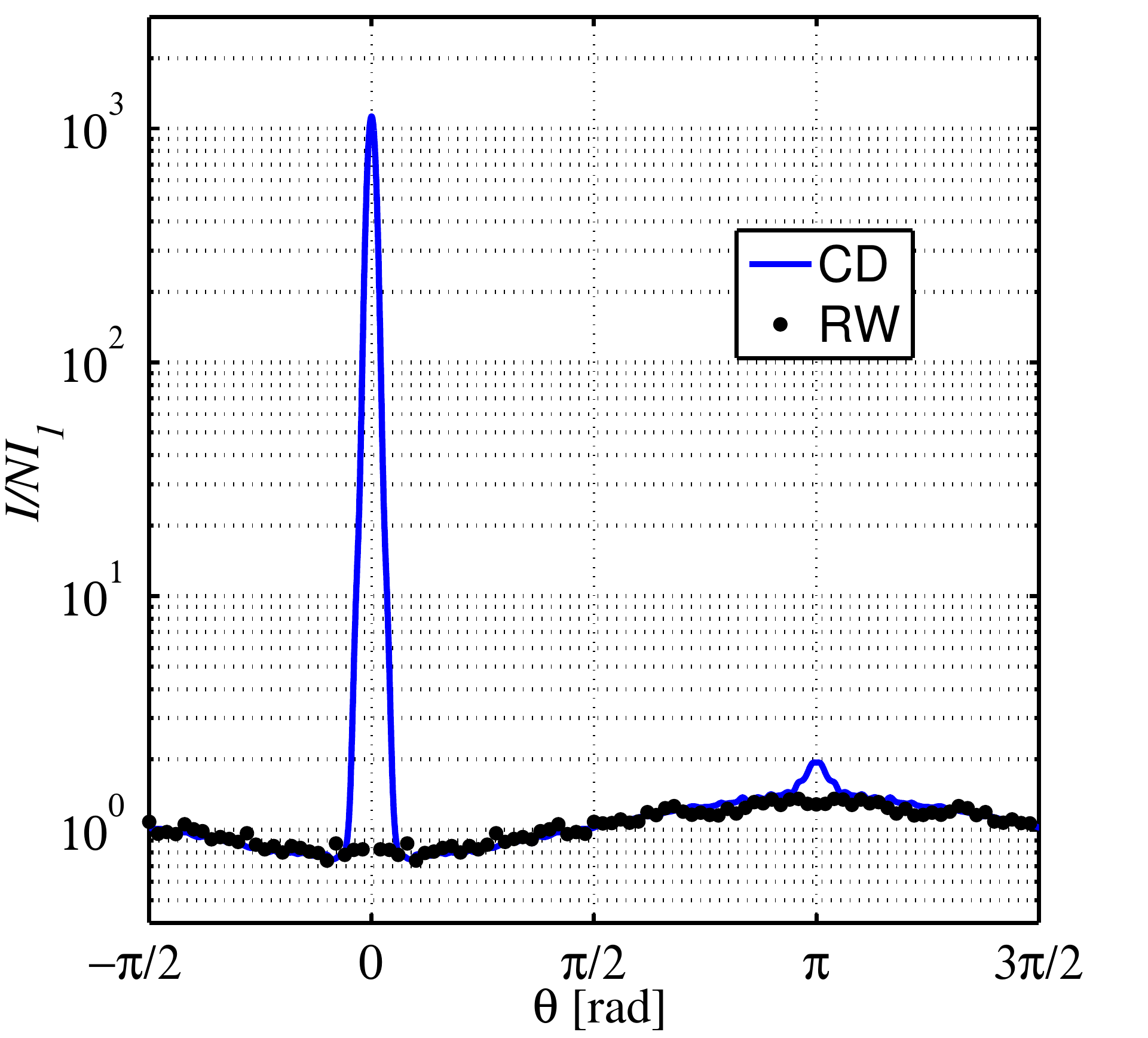}\caption{(color online). Emission diagram (log scale) normalized to the independent
N atom case, for the RW approach (black points) and the CD model (blue
line) with $b=8$. \label{fig:1 - Diagram}}
\end{figure}

The radiative transfer equation~\citep{PhysRevA.80.013831,Chandrasekhar}
is a very useful, and often used model, to describe the multiple scattering
regime that takes place in optically thick media. In such a system
one photon undergoes a large amount of independent scattering events
from randomly positioned particles and therefore interferences between
different paths are supposed to be smeared out by the disorder. Then,
the transport properties are mostly independent from the wave nature
of light and from the particular nature of the scatterers. Such an
incoherent model of multiple scattering of photons undergoing a random
walk inside the sample provides a good description of these systems,
in particular when considering configuration averaged signals. For
this incoherent random walk model, we perform a Monte Carlo simulation
where photons are isotropically scattered after a distance $l_{sc}(r)=1/[n(r)\sigma_{sc}]$,
where $\sigma_{sc}$ is the scattering cross section and $n(r)=n_{0}e^{-r^{2}/(2\sigma_{R}^{2})}$
the spatial density distribution of the cold atom cloud considered
in this work, well described by a spherical Gaussian distribution
of size $\sigma_{R}$ and center density $n_{0}$. The optical thickness
along one line of sight across the center of the cloud is defined
as $b=\int n(0,0,z)\sigma_{sc}dz=\sqrt{2\pi}n_{0}\sigma_{sc}\sigma_{R}=\sqrt{2\pi}\sigma_{R}/l_{sc}(r=0)$.
By integrating the extinction over the whole transverse size of the
cloud ($\int{dxdy~e^{-b(x,y)}}$), it is possible to obtain the total
extinction of an incident flux of photons, corresponding to a total
scattering cross section for this random walk model given by $\sigma_{RW}=2\pi\sigma_{R}^{2}~Ein(b)$,
where $Ein(b)$ is the entire function $Ein(b)=\int_{0}^{b}{(1-e^{-t})dt/t}\underset{b\rightarrow0}{\rightarrow}b$.
To compute the emission diagram, averaged over the azimuthal angle
$\phi$, $I(\theta)$ for this specific geometry, we simulated the
trajectories of $10^{5}$ photons arrived along the $z$ axis, over
an uniform disk of radius $R=4.36\sigma_{R}$ for a sample of $b=8$.
With $\approx28.2$\% of this ``plane'' wave diffused, this corresponds
to a scattering cross section for the gaussian cloud $\sigma_{RW}\approx16.8\sigma_{R}^{2}$,
very close to the analytical value $2\pi\sigma_{R}^{2}~Ein(8)\approx16.7\sigma_{R}^{2}$.
For each photon, we record the direction of emission obtaining the
angular emission diagram $I_{RW}(\theta)$ (see Fig.~\ref{fig:1 - Diagram}).

\section{Coupled Dipole Model}

We now compare this random walk approach to a model of coupled dipoles
that accounts for the interference between the radiation of all the
atoms. Even though light scattering in 3 dimensions requires to take
into account the polarization as well as the near-field dipole-dipole
coupling, a better comparison to our isotropic random walk approach
is obtained using a scalar model for the dipole-dipole coupling. In
the steady-state regime, the $N$ dipoles with amplitudes $\beta_{j}$
and position $\mathbf{r}_{j}$, illuminated by a quasi-resonant plane-wave
with wavevector $\mathbf{k}_{0}=k_{0}\hat{z}$, obey the equation~\citep{Courteille2010}:
\begin{eqnarray}
\left(i\delta-\frac{\Gamma}{2}\right)\beta_{j} & = & i\frac{\Omega_{0}}{2}e^{i\mathbf{k}_{0}\cdot\mathbf{r}_{j}}\label{eq:ManyBody}\\
 &  & +\frac{\Gamma}{2}\sum_{m\neq j}\beta_{m}\frac{\exp(ik_{0}|\mathbf{r}_{j}-\mathbf{r}_{m}|)}{ik_{0}|\mathbf{r}_{j}-\mathbf{r}_{m}|},\nonumber 
\end{eqnarray}
where $\omega_{a}$ is the resonance frequency, $\delta=\omega_{a}-\omega_{k_{0}}$
is the laser detuning, $\Gamma$ the atomic transition linewidth,
$\Omega_{0}=dE_{0}/\hbar$ the Rabi frequency and $d$ the dipole
matrix element. This many-body problem with interference has been
derived from a quantum formalism where a single photon is shared between
all atoms through a superposition of states~\citep{Courteille2010},
but also from a classical approach where the atoms are considered
as oscillators~\citep{Svidzinsky2010}. The relevant parameters to
describe light scattering in dilute clouds of two-level systems is
the resonant optical thickness of the cloud, which is given by $b_{0}=2N/(k_{0}\sigma_{R})^{2}$
with an on-resonant scattering cross section for a single atom given
in the scalar model by $\sigma_{sc}=\lambda^{2}/\pi$. The detuning
dependent optical thickness then reads: $b=b(\delta)=b_{0}/(1+4\delta^{2}/\Gamma^{2})$.
From \eqref{eq:ManyBody}, the far-field intensity in a direction
$\mathbf{k}$ and at a distance $r$ can be calculated using~\citep{JMO2013}:
\begin{equation}
4\pi r^{2}I(\hat{\mathbf{k}})=\hbar\omega_{k_{0}}\Gamma{\displaystyle \sum_{j,m=1}^{N}}\beta_{j}\beta_{m}^{*}e^{-ik_{0}\hat{\mathbf{k}}(\mathbf{r}_{j}-\mathbf{r}_{m})}.\label{eq:Emission_diagramMB}
\end{equation}
All numerical CD results data shown in this paper have been obtained
using an average over 20 different configurations of the atomic distribution.
We point out that the interferences are not only present in the emission
term \eqref{eq:Emission_diagramMB} (Rayleigh scattering), but already
in the steady-state value of the atomic dipoles~\eqref{eq:ManyBody}.
We can now compare the angular emission diagrams obtained from the
coherent CD equations to those from the incoherent RW model. In Fig.~\ref{fig:1 - Diagram}
we have normalized the emission diagram of the CD model to the emission
diagram of $N$ independent atoms (obtained by using a very large
sample size with vanishing optical thickness). The emission diagram
of the RW model has been normalized such that 
\begin{equation}
\frac{P_{CD}}{P_{RW}}=\frac{\sigma_{CD}}{\sigma_{RW}},
\end{equation}
where $P_{CD}$ (resp. $P_{RW}$) is the scattered power of the CD
(resp. RW) model for the same incident intensity. $P_{CD}$ and $\sigma_{CD}$
can be obtained from integration of the emission diagram $\int I(\hat{\mathbf{k}})d\hat{\mathbf{k}}$
or also from $\sigma_{CD}=-\frac{4\pi}{k_{0}^{2}}\frac{\Gamma}{\Omega_{0}}\textrm{Im}[\sum_{j}\beta_{j}e^{-i\mathbf{k_{0}}\cdot\mathbf{r}_{j}}]$
\cite{JMO2013}. The total cross section for the incoherent scattering
$\sigma_{RW}$ is obtained from the analytical expression given above.
For the parameters in Fig.~\ref{fig:1 - Diagram} ($b=8$, $\sigma_{R}=20/k_{0}$,
$\delta=0)$, we obtain $\sigma_{CD}=24.5\sigma_{R}^{2}$ and $\sigma_{RW}=16.7\sigma_{R}^{2}$.
The difference in these total scattering cross sections is mainly
explained by the additional strong forward lobe in the CD (see Fig.~\ref{fig:1 - Diagram}),
and can also be understood as the origin of the extinction paradox
\cite{JMO2013}.

As one can clearly see in Fig.~\ref{fig:1 - Diagram}, the RW model
in this multiple scattering regime is very close to the coherent CD
model, except for the coherent backscattering cone and the forward
lobe, with an angular width given by the inverse size of the sample.
The quantitative agreement between the RW and the CD model in all
angles except the forward and backward direction suggests that interferences
indeed appear washed out under multiple scattering conditions, as
naively expected. We note however that interferences might nevertheless
be relevant in multiple scattering, when spatially dense samples are
considered and one expects to approach the Anderson localization transition
or when going beyond the average emission diagram in steady state.

\section{Experimental Results}

The differences between coherent and incoherent scattering have been
addressed in the past, including experiments on cold atoms, with the
possibility to exploit the detuning as a relevant control parameter.
The enhanced backscattering cone, visible only in the CD model and
absent in the RW model, has been extensively studied more than 10
years ago \cite{PhysRevLett.83.5266}. The most important difference
between coherent and incoherent multiple scattering with a gaussian
shaped sample is the pronounced forward lobe. Detecting light scattered
in a direction close to the incident radiation is notoriously difficult
as most detecting schemes will be saturated by the large incident
radiation. An elegant technique has been used in \cite{Padmabandu:92}
but has not been implemented so far with atomic clouds. An alternative
approach to indirectly probe features of cooperative scattering by
laser cooled atoms has been used in the single scattering limit in
\cite{PhysRevLett.104.183602}, where the modification of the atomic
motion induced by light scattering has been used as a measure of cooperative
scattering. It is therefore interesting to investigate how the radiation
pressure force on the center of mass of the atoms extends into the
multiple scattering limit.

\begin{figure}
\centering{}\includegraphics[width=1\columnwidth]{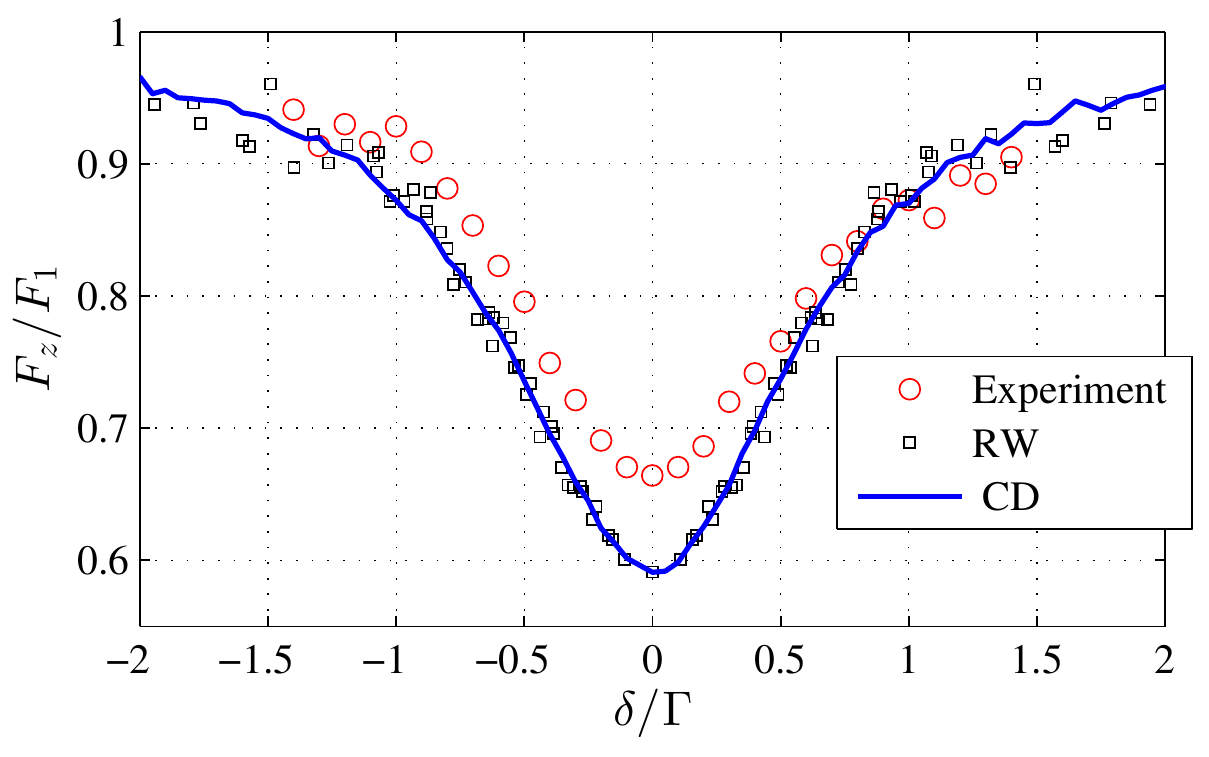}\caption{(color online) Normalized radiation pressure force acting on the center
of mass of the atomic cloud as a function of the laser detuning $\delta$ (in unit of $\Gamma$),
at constant saturation parameter $s(\delta)$. The experimental result
for $b_{0}=2.19$ (red circles) are compared to the coherent CD model
($b_{0}=2.19$, blue line) and incoherent RW model (black squares)
with corresponding optical thickness $b(\delta)$. \label{fig:2 - Force_delta}}
\end{figure}

We have therefore used the same experimental setup and protocol as
in \cite{PhysRevLett.104.183602}, and used values of the laser frequency
around the atomic resonance ($\delta\approx0$), thus entering the
multiple scattering limit. We apply the following experimental procedure
to probe the coherence of the multiple scattering regime in our cold
atom cloud. First, we load a magneto-optical trap (MOT) with $3\times10^{7}$
atoms of $^{87}$Rb in $50$ ms using the setup described in \citep{PhysRevLett.104.183602}.
We then apply a $50$ ms temporal dark MOT period where the intensity
of the repumping laser is reduced by a factor of $10$ and the cooling
laser is tuned to $-10\,\Gamma$ from the $F=2\rightarrow F'=3$ D$_{2}$
line. This allows to compress the cloud and to produce a smooth Gaussian
shaped distribution of atoms. To control the optical thickness at
the end of this dark MOT period, the repumper detuning is varied between
$-7\,\Gamma$ to $-2.5\,\Gamma$, keeping the desired amount of atoms
in the $F=2$ state without affecting size ($\sigma_{R}=270\,\mu$m,
$k\sigma_{R}\thickapprox2\times10^{3}$), shape and temperature ($\sim20\,\mu$K)
of the cloud. Here we focus on moderate values of optical thickness,
as this allowed for more systematic data without drifts of the relevant
parameters. We then switch off all laser beams and magnetic field
gradients, leaving the atoms in free fall. We then apply an horizontal,
circularly-polarized ``pushing'' beam, tuned close to the $F=2\rightarrow F'=3$
transition for $50\,\mu$s. The pushing beam has a waist $w_{0}=12$
mm and its carefully calibrated intensity is adjusted to have a saturation
parameter $s=8\times10^{-2}$. Each atom in the $F=2$ state scatters
on average $80$ photons. Such a small number of scattered photons
prevents from any depumping effect into the $F=1$ state during the
pushing process. After a time of flight expansion of $12$ ms, we
image the position of the atomic cloud via standard off resonant (detuning
$\delta\approx-2\,\Gamma$) absorption scheme. The absorption image
gives the position of the center of mass of the atomic cloud after
time of flight and thus the average radiation pressure force. Each
experimental point (see Fig.\ref{fig:2 - Force_delta}) is an average
over $10$ realizations. 
We normalize the measured average radiation pressure force by the
single atom force $F_{1}$, where $F_{1}$ is computed without any
adjustable parameter from the measured intensity of the pushing beam
(known within $5\,\%$ accuracy and taking into account losses by
the vacuum windows). The experimental value of the optical thickness
$b_{0}$ is obtained by standard absorption imaging, using a linearly
polarized laser beam. We note that such a transmission measurement
is best described by atoms in a statistical mixture of the Zeeman
sublevels in the ground state, with a corresponding average squared
Clebsch-Gordon coefficient of $(2F'+1)/3(2F+1)=7/15$. The effective
resonant optical thickness for a spherical gaussian cloud of atoms
distributed in a statistical mixture of the Zeeman sublevels is thus
given by $b_{0}=\frac{2F'+1}{3(2F+1)}\frac{3N}{k\sigma_{R}^{2}}$,
in contrast to a situation where all atoms would be pumped with a
circularly polarized laser beam into the stretched state $|F=2,m_{F}=+2\rangle$.

In Fig. \ref{fig:2 - Force_delta}, we show the experimental result
of the intrinsic radiation pressure force, proportional to the displacement
of the center of mass of the atomic cloud (red spheres) as a function
of the pushing beam detuning $\delta$, where we keep the saturation
parameter $s=s_{0}/(1+4\delta^{2}/\Gamma^{2})$ constant. We clearly
see an important reduction of the intrinsic radiation pressure force
around the resonance. Note that with this experimental protocol, the
mass of the atomic sample to be displaced as well as the single atom
response are kept constant, allowing to highlight the collective behavior.

We now turn to the comparison between the experimental data and our
coherent and incoherent models. Momentum conservation arguments allow
to directly connect the far field emission diagram of a sample to
the momentum transfer to the center of mass of the sample. We stress
that it is important not only to consider the shape of the emission
diagram, but also the total scattered power, which can depend on the
shape and opacity of the sample. Defining the intrinsic radiation
pressure force along $\hat{e}_{z}$ by $F_{z}=(1/N)\sum_{j}F_{j}$,
where $F_{j}$ is the force acting on atom $j$ \cite{Courteille2010},
we obtain the following relation 
\begin{equation}
\dfrac{F_{z}}{F_{1}}=\dfrac{\sigma_{tot}}{N\sigma_{1}}\langle1-\cos(\theta)\rangle,\label{force1}
\end{equation}
where $\cos\theta=\hat{\mathbf{k}}\cdot\hat{z}$ is the angle of the
direction of emitted photon with the laser axis. We note that this
relation holds both for the incoherent RW and coherent CD model. In
(\ref{force1}), $F_{1}$ is the radiation pressure force for a single
atom, and $\sigma_{tot}$ is the total scattering of the sample, which
is different in the RW and the CD model. One can see from (\ref{force1})
that the emission diagram associated with the total power scattered
from the incident beam allows to predict the intrinsic force, proportional
to the acceleration of the center of mass of the sample. The numerical
results of our CD solution (blue line) also show a strong reduction
of the force. To come forward with a simple interpretation of these
results, one can associate the term corresponding to $\dfrac{\sigma_{tot}}{N\sigma_{1}}$
as a ``shadow\textquotedbl{} effect, where that part of the force
is obtained with only taking into account the attenuation of the incident
laser beam. The recoil due to the rescattered photons is then properly
taken into account by the last term of (\ref{force1}): $-\dfrac{\sigma_{tot}}{N\sigma_{1}}\langle\cos(\theta)\rangle$,
related to the emission diagram of the photons. As shown in Fig. \ref{fig:2 - Force_delta}
the experimental result compares reasonably well both CD and RW model
when using the optical thickness evaluated for a statistical mixture
of Zeeman sublevels, assumed to be a good approximation under multiple
scattering conditions, when the local polarization is well approximated
by a field with random polarization.

\begin{figure}
\centering{} \includegraphics[width=1.1\columnwidth]{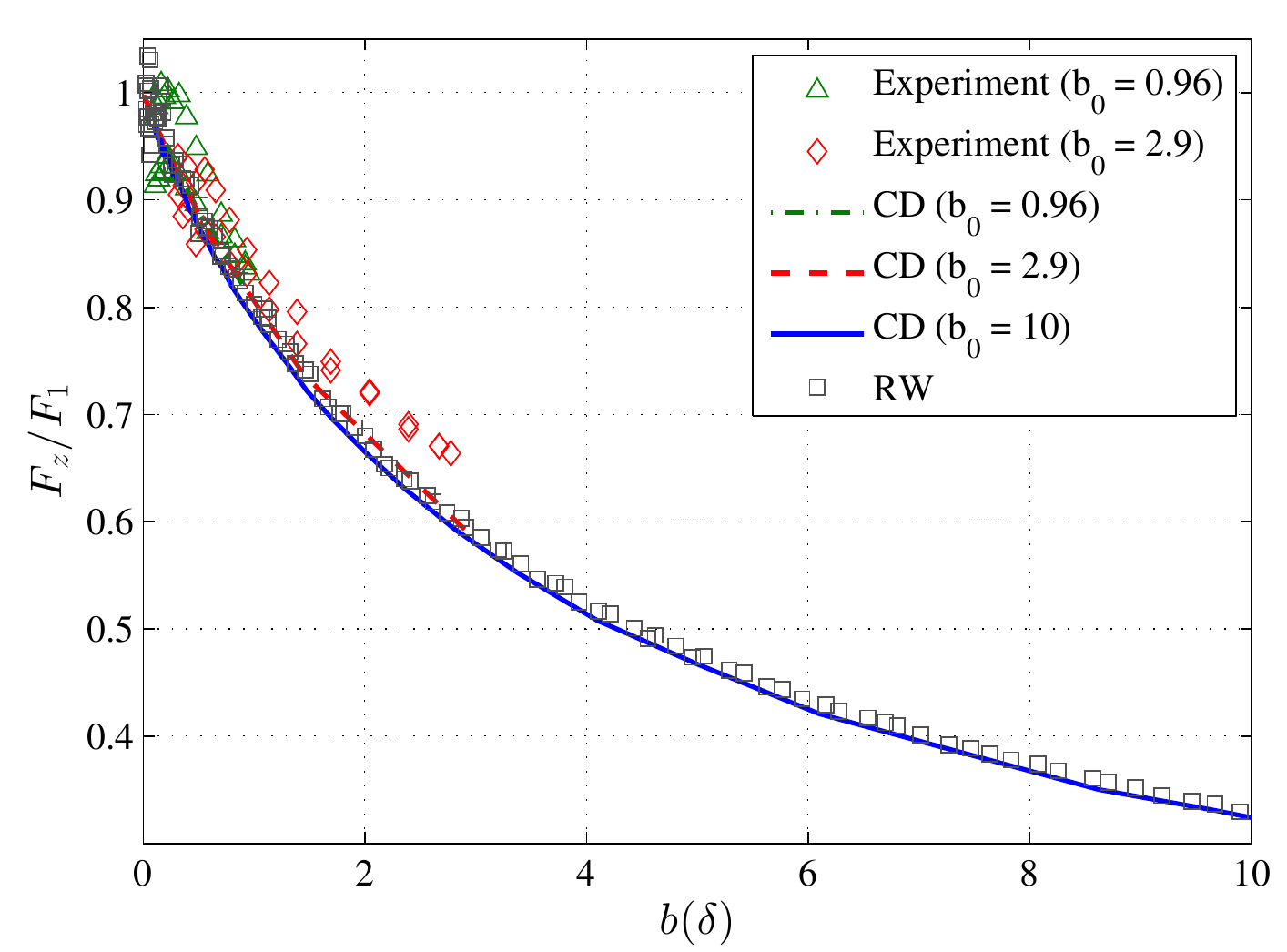}
\caption{(color online) Normalized radiation pressure force as a function of
the optical thickness. The experimental data for $b_{0}=0.96$ (green triangles) and $b_{0}=2.9$ (red diamonds) are compared to the results of  the CD simulations, obtained for $b_{0}=0.96$ (green dash-dotted line), $b_{0}=2.9$ (red dotted line) and $b_0=10$ (blue plain line), and of the RW model (gray squares). For the experimental data and the CD model, $b(\delta)$ is varied at fixed $b_{0}$ by changing the detuning. The CD data are superimposed as $b(\delta)$ appears as a universal parameter, and fluctuations are very small.}
\label{Fig3} 
\end{figure}

\section{Discussion}

In order to probe coherence effects under multiple scattering conditions,
it is important to check to what extent an incoherent model also explains
similar features. When looking at the prediction of the incoherent
model, we find that there is no significant difference in the regime
of parameters studied. This very close match between the incoherent
and coherent model can be explained by the fact that the emission
diagram only differ at two angles. In backward direction, the coherent
backscattering cone should result in a small increase of the intrinsic
radiation pressure force in the coherent model compared to an incoherent
model. However, the enhancement factor in backward direction is less
than 2 and the angular range of enhanced backscattering is very small
for a dilute sample of atoms. On the other hand, the more important
emission intensity in the forward lobe does not result in significant
change in the intrinsic radiation pressure force, as in this direction
($\theta\approx0$) the momentum transfer to the atoms is vanishing:
$\langle1-\cos(\theta)\rangle\approx0$. It thus turns out that under
these conditions, the intrinsic radiation pressure force is not a
good measure to detect differences between coherent and incoherent
scattering. We further investigate the range of validity of the RW
model in Fig. \ref{Fig3}, where we plot the intrinsic force extracted
from the experimental data for various $b_{0}$, the RW simulation
and the CD simulation as a function of $b(\delta)$. The RW simulation
is by definition a function of $b(\delta)$ only. A close inspection
of the experimental and CD result shows small deviations from the
RW model, including a red/blue asymmetry, visible at small $b(\delta)$. The origin of this asymmetry remains unknown. However, we observed it on CD data  for much higher $b_0$ than experimentally measured.
When looking for coherence effects, one does indeed expect to find
cooperative signatures for large $b_{0}$ and large detunings, where
a mean field approach is assumed to be valid \cite{PhysRevLett.104.183602}.
Another feature not included in the RW model are Mie resonances expected
in the CD model for large $b_{0}$ \cite{BachelardMie}. We also note
a small deviation of the experimental results compared to both RW
and CD model close to resonance ($\delta=0$) (see Fig. \ref{fig:2 - Force_delta}
and Fig. \ref{Fig3}). This difference would correspond to a slightly
smaller atom-light coupling than obtained from a statistical mixture
of atoms in the Zeeman sublevels, which would however be rather surprising.

\begin{figure}[t!]
\centering{}\includegraphics[width=1\columnwidth]{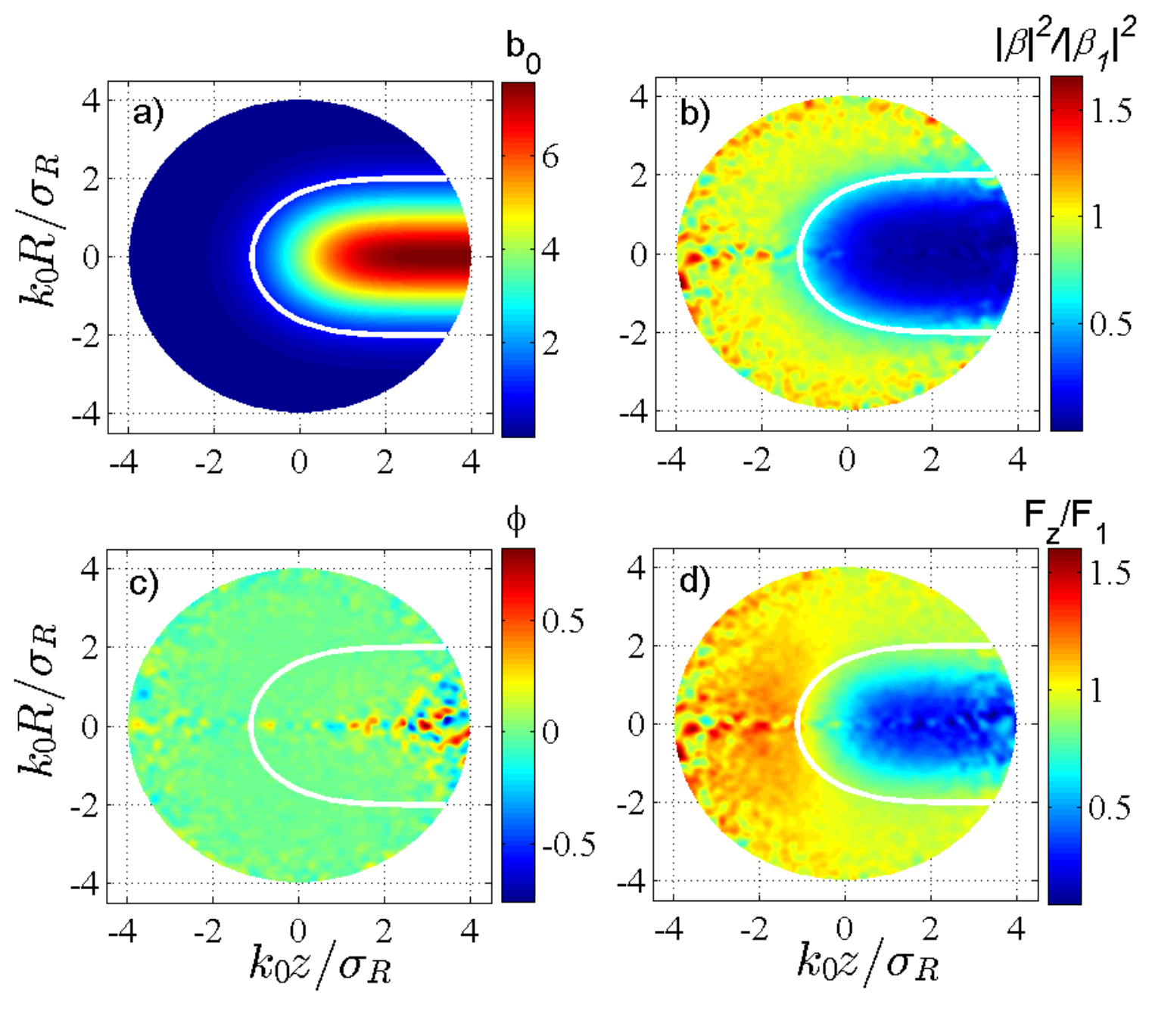}\caption{(color online). a) Local optical thickness $b_{0}$ seen by the laser
field propagating along the direction $\hat{e}_{z}$. The white line
represents the multiple scattering limit $b_{0}=1$. b) Excitation
amplitude of the atomic dipole compared to the single atom limit $\left|\beta_{j}\right|^{2}/\left|\beta_{1}\right|^{2}$.
c) Phase of the atomic dipole $\phi_{j}$. $\phi=0$ corresponds to
the laser field phase corrected from its propagation along $\hat{e}_{z}$.
d) Local force compared to the single atom case $F_{z}/F_{1}$. All
data are computed for a gaussian sphere with $b_{0}=8$. \label{fig:4 -Phases}}
\end{figure}

Finally, we plotted on Fig. \ref{fig:4 -Phases} various data extracted
from our CD simulation using the same gaussian sphere that gives the
emission diagram plotted on Fig. \ref{fig:1 - Diagram}. Fig. \ref{fig:4 -Phases}a
represents the attenuation $e^{-b_{0}}$ of the incident laser field
at the position $\hat{e}_{z}$ and along the transverse direction
$\hat{e}_{r}$. Figure \ref{fig:4 -Phases}b represents the dipole
excitation relative to the single atom limit $\left|\beta_{j}\right|^{2}/\left|\beta_{1}\right|^{2}$.
For an integrated opacity $b_{0}>1$ (white line), one expects the
atomic dipoles to be less excited by the incident laser field. Those
atoms in the shadow are indeed significantly less excited. Figure
\ref{fig:4 -Phases}c represents the phase of the atomic dipoles.
Most of the atomic cloud has a phase $\phi=0$ corresponding to the
laser field. In other words, almost the entire cloud is synchronized
with the incident laser field even for $b_{0}>1$. This is somewhat
in contradiction to what one would expect from a RW hypothesis, where
in multiple scattering regime the phase is randomized after few scattering
events and interference effects are smeared out by the disorder. This
result is however in good agreement with the Ewald-Oseen extinction
theorem \cite{ballenegger:599} which specifies that the radiation
from the atoms exactly cancels the electromagnetic field of the incident
laser beam and replaces it by a field with a speed $c/n$ where $n$
is the index of refraction of the medium. As shown in \citep{ballenegger:599},
not only the surface layer of the atomic cloud is synchronized but
the entire cloud participates to the cancellation of the incident
field. This is in contrast with the common belief that optically thick
samples are well described by the RW approach where the phase of a
photon is randomized after few diffusion events, but is in agreement
with related work \citep{Yang:99, Mishchenko} where a small persisting coherent
component in forward scattered field has been reported.
Fig. \ref{fig:4 -Phases}d represents the local force. Although the
atoms at the entrance of the cloud undergo a force larger than the
single-atom force, those in the shadow are subjected to a lower force,
resulting in a total force below the single-atom one, in agreement
with the measurement reported in Fig.~\ref{Fig3}.

\section{Conclusion}

In summary, we have compared an incoherent and a coherent model for
multiple scattering. The most prominent differences are the well known
coherent backscattering cone and an important forward lobe. We have
compared the predictions of these two models to the experimental result
of the intrinsic radiation pressure force acting on the center of
mass of the atomic cloud and found that this force is not a good candidate
to detect coherence effects in multiple scattering. Using an effective
coupling strength for the atom light coupling in multiple scattering,
we find a satisfactory quantitative agreement between the experiment
and the numerical model. It would be interesting to study how these
results compare to previous theoretical and experimental results,
where a mean field approach in terms of single photon superradiance
has been used. A precise experimental study of how multiple scattering
sets in and to what extend a RW model can explain the whole range
from large to small optical thickness is a complex task, as one needs
to take into account the various Zeeman sublevels of the Rubidium
atoms. We expect that alternative observables beyond average values
of the center of mass displacement of the cloud, such as correlations
in the scattered light or heating of the cloud, might be good candidates
to look for features of cooperativity in light scattering of light
by cold atoms.

\acknowledgments{R.K. acknowledges valuable discussions with D.
Delande. We acknowledge financial support from IRSES project COSCALI,
from USP/COFECUB (projet Uc Ph 123/11) and from GDRI nanomagnetism,
spin electroncics, quantum optics and quantum technologies. M. T.
R. is supported by an Averroès exchange program. R. B. acknowledges
support from the Fundação de Amparo à Pesquisa do Estado de São Paulo
(FAPESP).}

 \bibliographystyle{apsrev}
\bibliography{CMS_Biblio}

\begin{thebibliography}{30}
\expandafter\ifx\csname natexlab\endcsname\relax\def\natexlab#1{#1}\fi
\expandafter\ifx\csname bibnamefont\endcsname\relax
  \def\bibnamefont#1{#1}\fi
\expandafter\ifx\csname bibfnamefont\endcsname\relax
  \def\bibfnamefont#1{#1}\fi
\expandafter\ifx\csname citenamefont\endcsname\relax
  \def\citenamefont#1{#1}\fi
\expandafter\ifx\csname url\endcsname\relax
  \def\url#1{\texttt{#1}}\fi
\expandafter\ifx\csname urlprefix\endcsname\relax\def\urlprefix{URL }\fi
\providecommand{\bibinfo}[2]{#2}
\providecommand{\eprint}[2][]{\url{#2}}

\bibitem[{\citenamefont{Albada and Lagendijk}(1985)}]{PhysRevLett.55.2692}
\bibinfo{author}{\bibfnamefont{M.~P.~V.} \bibnamefont{Albada}}
  \bibnamefont{and}
  \bibinfo{author}{\bibfnamefont{A.}~\bibnamefont{Lagendijk}},
  \bibinfo{journal}{Phys. Rev. Lett.} \textbf{\bibinfo{volume}{55}},
  \bibinfo{pages}{2692} (\bibinfo{year}{1985}),
  \urlprefix\url{http://link.aps.org/doi/10.1103/PhysRevLett.55.2692}.

\bibitem[{\citenamefont{Wolf and Maret}(1985)}]{PhysRevLett.55.2696}
\bibinfo{author}{\bibfnamefont{P.-E.} \bibnamefont{Wolf}} \bibnamefont{and}
  \bibinfo{author}{\bibfnamefont{G.}~\bibnamefont{Maret}},
  \bibinfo{journal}{Phys. Rev. Lett.} \textbf{\bibinfo{volume}{55}},
  \bibinfo{pages}{2696} (\bibinfo{year}{1985}),
  \urlprefix\url{http://link.aps.org/doi/10.1103/PhysRevLett.55.2696}.

\bibitem[{\citenamefont{Wiersma et~al.}(1997)\citenamefont{Wiersma, Bartolini,
  Lagendijk, and Righini}}]{WiesmaLocLight97}
\bibinfo{author}{\bibfnamefont{D.~S.} \bibnamefont{Wiersma}},
  \bibinfo{author}{\bibfnamefont{P.}~\bibnamefont{Bartolini}},
  \bibinfo{author}{\bibfnamefont{A.}~\bibnamefont{Lagendijk}},
  \bibnamefont{and} \bibinfo{author}{\bibfnamefont{R.}~\bibnamefont{Righini}},
  \bibinfo{journal}{Nature} \textbf{\bibinfo{volume}{390}},
  \bibinfo{pages}{671} (\bibinfo{year}{1997}).

\bibitem[{\citenamefont{F.~Scheffold and Maret}(1999)}]{Maret1999}
\bibinfo{author}{\bibfnamefont{R.~T.} \bibnamefont{F.~Scheffold},
  \bibfnamefont{R.~Lenke}} \bibnamefont{and}
  \bibinfo{author}{\bibfnamefont{G.}~\bibnamefont{Maret}},
  \bibinfo{journal}{Nature} \textbf{\bibinfo{volume}{398}},
  \bibinfo{pages}{206} (\bibinfo{year}{1999}).

\bibitem[{\citenamefont{Wiersma et~al.}(1999)\citenamefont{Wiersma, Bartolini,
  Lagendijk, and Righini}}]{Lagendijk1999}
\bibinfo{author}{\bibfnamefont{D.~S.} \bibnamefont{Wiersma}},
  \bibinfo{author}{\bibfnamefont{P.}~\bibnamefont{Bartolini}},
  \bibinfo{author}{\bibfnamefont{A.}~\bibnamefont{Lagendijk}},
  \bibnamefont{and} \bibinfo{author}{\bibfnamefont{R.}~\bibnamefont{Righini}},
  \bibinfo{journal}{Nature} \textbf{\bibinfo{volume}{398}},
  \bibinfo{pages}{207} (\bibinfo{year}{1999}).

\bibitem[{\citenamefont{Lagendijk et~al.}(2009)\citenamefont{Lagendijk, van
  Tiggelen, and Wiersma}}]{PhysicsToday}
\bibinfo{author}{\bibfnamefont{A.}~\bibnamefont{Lagendijk}},
  \bibinfo{author}{\bibfnamefont{B.}~\bibnamefont{van Tiggelen}},
  \bibnamefont{and} \bibinfo{author}{\bibfnamefont{D.~S.}
  \bibnamefont{Wiersma}}, \bibinfo{journal}{Physics Today}
  \textbf{\bibinfo{volume}{62}}, \bibinfo{pages}{24} (\bibinfo{year}{2009}),
  \urlprefix\url{http://dx.doi.org/10.1063/1.3206091}.

\bibitem[{\citenamefont{Sperling et~al.}(2012)\citenamefont{Sperling, Buehrer,
  Aegerter, and Maret}}]{Maret2013a}
\bibinfo{author}{\bibfnamefont{T.}~\bibnamefont{Sperling}},
  \bibinfo{author}{\bibfnamefont{W.}~\bibnamefont{Buehrer}},
  \bibinfo{author}{\bibfnamefont{C.~M.} \bibnamefont{Aegerter}},
  \bibnamefont{and} \bibinfo{author}{\bibfnamefont{G.}~\bibnamefont{Maret}},
  \bibinfo{journal}{Nature Photonics} \textbf{\bibinfo{volume}{7}},
  \bibinfo{pages}{48} (\bibinfo{year}{2012}).

\bibitem[{\citenamefont{Scheffold and Wiersma}(2013)}]{scheffoldinelastic2013}
\bibinfo{author}{\bibfnamefont{F.}~\bibnamefont{Scheffold}} \bibnamefont{and}
  \bibinfo{author}{\bibfnamefont{D.}~\bibnamefont{Wiersma}},
  \bibinfo{journal}{Nature Photonics} \textbf{\bibinfo{volume}{7}},
  \bibinfo{pages}{934} (\bibinfo{year}{2013}),
  \urlprefix\url{http://dx.doi.org/10.1038/nphoton.2013.210}.

\bibitem[{\citenamefont{Maret et~al.}(2013)\citenamefont{Maret, Sperling,
  B{\"u}hrer, Lubatsch, Frank, and Aegerter}}]{maret2013inelastic}
\bibinfo{author}{\bibfnamefont{G.}~\bibnamefont{Maret}},
  \bibinfo{author}{\bibfnamefont{T.}~\bibnamefont{Sperling}},
  \bibinfo{author}{\bibfnamefont{W.}~\bibnamefont{B{\"u}hrer}},
  \bibinfo{author}{\bibfnamefont{A.}~\bibnamefont{Lubatsch}},
  \bibinfo{author}{\bibfnamefont{R.}~\bibnamefont{Frank}}, \bibnamefont{and}
  \bibinfo{author}{\bibfnamefont{C.~M.} \bibnamefont{Aegerter}},
  \bibinfo{journal}{Nature Photonics} \textbf{\bibinfo{volume}{7}},
  \bibinfo{pages}{934} (\bibinfo{year}{2013}).

\bibitem[{\citenamefont{van Tiggelen et~al.}(1990)\citenamefont{van Tiggelen,
  Lagendijk, and Tip}}]{CFS_BvT}
\bibinfo{author}{\bibfnamefont{B.~A.} \bibnamefont{van Tiggelen}},
  \bibinfo{author}{\bibfnamefont{A.}~\bibnamefont{Lagendijk}},
  \bibnamefont{and} \bibinfo{author}{\bibfnamefont{A.}~\bibnamefont{Tip}},
  \bibinfo{journal}{Journal of Physics: Condensed Matter}
  \textbf{\bibinfo{volume}{2}}, \bibinfo{pages}{7653} (\bibinfo{year}{1990}),
  \urlprefix\url{http://stacks.iop.org/0953-8984/2/i=37/a=010}.

\bibitem[{\citenamefont{Karpiuk et~al.}(2012)\citenamefont{Karpiuk, Cherroret,
  Lee, Gr\'emaud, M\"uller, and Miniatura}}]{CFS_Miniat}
\bibinfo{author}{\bibfnamefont{T.}~\bibnamefont{Karpiuk}},
  \bibinfo{author}{\bibfnamefont{N.}~\bibnamefont{Cherroret}},
  \bibinfo{author}{\bibfnamefont{K.~L.} \bibnamefont{Lee}},
  \bibinfo{author}{\bibfnamefont{B.}~\bibnamefont{Gr\'emaud}},
  \bibinfo{author}{\bibfnamefont{C.~A.} \bibnamefont{M\"uller}},
  \bibnamefont{and}
  \bibinfo{author}{\bibfnamefont{C.}~\bibnamefont{Miniatura}},
  \bibinfo{journal}{Phys. Rev. Lett.} \textbf{\bibinfo{volume}{109}},
  \bibinfo{pages}{190601} (\bibinfo{year}{2012}),
  \urlprefix\url{http://link.aps.org/doi/10.1103/PhysRevLett.109.190601}.

\bibitem[{\citenamefont{Scheffold and Maret}(1998)}]{PhysRevLett.81.5800}
\bibinfo{author}{\bibfnamefont{F.}~\bibnamefont{Scheffold}} \bibnamefont{and}
  \bibinfo{author}{\bibfnamefont{G.}~\bibnamefont{Maret}},
  \bibinfo{journal}{Phys. Rev. Lett.} \textbf{\bibinfo{volume}{81}},
  \bibinfo{pages}{5800} (\bibinfo{year}{1998}),
  \urlprefix\url{http://link.aps.org/doi/10.1103/PhysRevLett.81.5800}.

\bibitem[{\citenamefont{Krachmalnicoff
  et~al.}(2010)\citenamefont{Krachmalnicoff, Castani{\'e}, De~Wilde, and
  Carminati}}]{Carminati}
\bibinfo{author}{\bibfnamefont{V.}~\bibnamefont{Krachmalnicoff}},
  \bibinfo{author}{\bibfnamefont{E.}~\bibnamefont{Castani{\'e}}},
  \bibinfo{author}{\bibfnamefont{Y.}~\bibnamefont{De~Wilde}}, \bibnamefont{and}
  \bibinfo{author}{\bibfnamefont{R.}~\bibnamefont{Carminati}},
  \bibinfo{journal}{Phys. Rev. Lett} \textbf{\bibinfo{volume}{105}},
  \bibinfo{pages}{183901} (\bibinfo{year}{2010}).

\bibitem[{\citenamefont{Liu and Lalanne}(2008)}]{Lalanne}
\bibinfo{author}{\bibfnamefont{H.}~\bibnamefont{Liu}} \bibnamefont{and}
  \bibinfo{author}{\bibfnamefont{P.}~\bibnamefont{Lalanne}},
  \bibinfo{journal}{Nature} \textbf{\bibinfo{volume}{452}},
  \bibinfo{pages}{728} (\bibinfo{year}{2008}).

\bibitem[{\citenamefont{Dicke}(1954)}]{Dicke1954}
\bibinfo{author}{\bibfnamefont{R.~H.} \bibnamefont{Dicke}},
  \bibinfo{journal}{Phys. Rev.} \textbf{\bibinfo{volume}{93}},
  \bibinfo{pages}{99} (\bibinfo{year}{1954}).

\bibitem[{\citenamefont{Courteille et~al.}(2010)\citenamefont{Courteille, Bux,
  Lucioni, Lauber, Bienaim\'e, Kaiser, and Piovella}}]{Courteille2010}
\bibinfo{author}{\bibfnamefont{P.~W.} \bibnamefont{Courteille}},
  \bibinfo{author}{\bibfnamefont{S.}~\bibnamefont{Bux}},
  \bibinfo{author}{\bibfnamefont{E.}~\bibnamefont{Lucioni}},
  \bibinfo{author}{\bibfnamefont{K.}~\bibnamefont{Lauber}},
  \bibinfo{author}{\bibfnamefont{T.}~\bibnamefont{Bienaim\'e}},
  \bibinfo{author}{\bibfnamefont{R.}~\bibnamefont{Kaiser}}, \bibnamefont{and}
  \bibinfo{author}{\bibfnamefont{N.}~\bibnamefont{Piovella}},
  \bibinfo{journal}{The European Physical Journal D}
  \textbf{\bibinfo{volume}{58}}, \bibinfo{pages}{69} (\bibinfo{year}{2010}),
  ISSN \bibinfo{issn}{1434-6060},
  \urlprefix\url{http://dx.doi.org/10.1140/epjd/e2010-00095-6}.

\bibitem[{\citenamefont{Hald et~al.}(1999)\citenamefont{Hald, S\o{}rensen,
  Schori, and Polzik}}]{PhysRevLett.83.1319}
\bibinfo{author}{\bibfnamefont{J.}~\bibnamefont{Hald}},
  \bibinfo{author}{\bibfnamefont{J.~L.} \bibnamefont{S\o{}rensen}},
  \bibinfo{author}{\bibfnamefont{C.}~\bibnamefont{Schori}}, \bibnamefont{and}
  \bibinfo{author}{\bibfnamefont{E.~S.} \bibnamefont{Polzik}},
  \bibinfo{journal}{Phys. Rev. Lett.} \textbf{\bibinfo{volume}{83}},
  \bibinfo{pages}{1319} (\bibinfo{year}{1999}),
  \urlprefix\url{http://link.aps.org/doi/10.1103/PhysRevLett.83.1319}.

\bibitem[{\citenamefont{Bloom et~al.}(2014)\citenamefont{Bloom, Nicholson,
  Williams, Campbell, Bishof, Zhang, Zhang, Bromley, and Ye}}]{Ye_Science2014}
\bibinfo{author}{\bibfnamefont{B.~J.} \bibnamefont{Bloom}},
  \bibinfo{author}{\bibfnamefont{T.~L.} \bibnamefont{Nicholson}},
  \bibinfo{author}{\bibfnamefont{J.~R.} \bibnamefont{Williams}},
  \bibinfo{author}{\bibfnamefont{S.~L.} \bibnamefont{Campbell}},
  \bibinfo{author}{\bibfnamefont{M.}~\bibnamefont{Bishof}},
  \bibinfo{author}{\bibfnamefont{X.}~\bibnamefont{Zhang}},
  \bibinfo{author}{\bibfnamefont{W.}~\bibnamefont{Zhang}},
  \bibinfo{author}{\bibfnamefont{S.~L.} \bibnamefont{Bromley}},
  \bibnamefont{and} \bibinfo{author}{\bibfnamefont{J.}~\bibnamefont{Ye}},
  \bibinfo{journal}{Nature}  (\bibinfo{year}{2014}),
  \urlprefix\url{http://www.nature.com/doifinder/10.1038/nature12941}.

\bibitem[{\citenamefont{de~Vries et~al.}(1998)\citenamefont{de~Vries, van
  Coevorden, and Lagendijk}}]{RevModPhys.70.447}
\bibinfo{author}{\bibfnamefont{P.}~\bibnamefont{de~Vries}},
  \bibinfo{author}{\bibfnamefont{D.~V.} \bibnamefont{van Coevorden}},
  \bibnamefont{and}
  \bibinfo{author}{\bibfnamefont{A.}~\bibnamefont{Lagendijk}},
  \bibinfo{journal}{Rev. Mod. Phys.} \textbf{\bibinfo{volume}{70}},
  \bibinfo{pages}{447} (\bibinfo{year}{1998}),
  \urlprefix\url{http://link.aps.org/doi/10.1103/RevModPhys.70.447}.

\bibitem[{\citenamefont{Pierrat et~al.}(2009)\citenamefont{Pierrat, Gr\'emaud,
  and Delande}}]{PhysRevA.80.013831}
\bibinfo{author}{\bibfnamefont{R.}~\bibnamefont{Pierrat}},
  \bibinfo{author}{\bibfnamefont{B.}~\bibnamefont{Gr\'emaud}},
  \bibnamefont{and} \bibinfo{author}{\bibfnamefont{D.}~\bibnamefont{Delande}},
  \bibinfo{journal}{Phys. Rev. A} \textbf{\bibinfo{volume}{80}},
  \bibinfo{pages}{013831} (\bibinfo{year}{2009}),
  \urlprefix\url{http://link.aps.org/doi/10.1103/PhysRevA.80.013831}.

\bibitem[{\citenamefont{Chandrasekhar}(1960)}]{Chandrasekhar}
\bibinfo{author}{\bibfnamefont{S.}~\bibnamefont{Chandrasekhar}},
  \emph{\bibinfo{title}{Radiative Transfer}} (\bibinfo{publisher}{Dover, New
  York}, \bibinfo{year}{1960}).

\bibitem[{\citenamefont{Svidzinsky et~al.}(2010)\citenamefont{Svidzinsky,
  Chang, and Scully}}]{Svidzinsky2010}
\bibinfo{author}{\bibfnamefont{A.~A.} \bibnamefont{Svidzinsky}},
  \bibinfo{author}{\bibfnamefont{J.}~\bibnamefont{Chang}}, \bibnamefont{and}
  \bibinfo{author}{\bibfnamefont{M.~O.} \bibnamefont{Scully}},
  \bibinfo{journal}{Phys. Rev. A} \textbf{\bibinfo{volume}{81}},
  \bibinfo{pages}{053821} (\bibinfo{year}{2010}).

\bibitem[{\citenamefont{Bienaimé et~al.}(2014)\citenamefont{Bienaimé,
  Bachelard, Chabé, Rouabah, Bellando, Courteille, Piovella, and
  Kaiser}}]{JMO2013}
\bibinfo{author}{\bibfnamefont{T.}~\bibnamefont{Bienaimé}},
  \bibinfo{author}{\bibfnamefont{R.}~\bibnamefont{Bachelard}},
  \bibinfo{author}{\bibfnamefont{J.}~\bibnamefont{Chabé}},
  \bibinfo{author}{\bibfnamefont{M.~T.} \bibnamefont{Rouabah}},
  \bibinfo{author}{\bibfnamefont{L.}~\bibnamefont{Bellando}},
  \bibinfo{author}{\bibfnamefont{P.~W.} \bibnamefont{Courteille}},
  \bibinfo{author}{\bibfnamefont{N.}~\bibnamefont{Piovella}}, \bibnamefont{and}
  \bibinfo{author}{\bibfnamefont{R.}~\bibnamefont{Kaiser}},
  \bibinfo{journal}{J. Mod. Opt.} \textbf{\bibinfo{volume}{61}},
  \bibinfo{pages}{18} (\bibinfo{year}{2014}).

\bibitem[{\citenamefont{Labeyrie et~al.}(1999)\citenamefont{Labeyrie,
  de~Tomasi, Bernard, M\"uller, Miniatura, and Kaiser}}]{PhysRevLett.83.5266}
\bibinfo{author}{\bibfnamefont{G.}~\bibnamefont{Labeyrie}},
  \bibinfo{author}{\bibfnamefont{F.}~\bibnamefont{de~Tomasi}},
  \bibinfo{author}{\bibfnamefont{J.-C.} \bibnamefont{Bernard}},
  \bibinfo{author}{\bibfnamefont{C.~A.} \bibnamefont{M\"uller}},
  \bibinfo{author}{\bibfnamefont{C.}~\bibnamefont{Miniatura}},
  \bibnamefont{and} \bibinfo{author}{\bibfnamefont{R.}~\bibnamefont{Kaiser}},
  \bibinfo{journal}{Phys. Rev. Lett.} \textbf{\bibinfo{volume}{83}},
  \bibinfo{pages}{5266} (\bibinfo{year}{1999}),
  \urlprefix\url{http://link.aps.org/doi/10.1103/PhysRevLett.83.5266}.

\bibitem[{\citenamefont{Padmabandu et~al.}(1992)\citenamefont{Padmabandu, Oh,
  and Fry}}]{Padmabandu:92}
\bibinfo{author}{\bibfnamefont{G.~G.} \bibnamefont{Padmabandu}},
  \bibinfo{author}{\bibfnamefont{C.}~\bibnamefont{Oh}}, \bibnamefont{and}
  \bibinfo{author}{\bibfnamefont{E.~S.} \bibnamefont{Fry}},
  \bibinfo{journal}{Opt. Lett.} \textbf{\bibinfo{volume}{17}},
  \bibinfo{pages}{169} (\bibinfo{year}{1992}),
  \urlprefix\url{http://ol.osa.org/abstract.cfm?URI=ol-17-3-169}.

\bibitem[{\citenamefont{Bienaim\'e et~al.}(2010)\citenamefont{Bienaim\'e, Bux,
  Lucioni, Courteille, Piovella, and Kaiser}}]{PhysRevLett.104.183602}
\bibinfo{author}{\bibfnamefont{T.}~\bibnamefont{Bienaim\'e}},
  \bibinfo{author}{\bibfnamefont{S.}~\bibnamefont{Bux}},
  \bibinfo{author}{\bibfnamefont{E.}~\bibnamefont{Lucioni}},
  \bibinfo{author}{\bibfnamefont{P.~W.} \bibnamefont{Courteille}},
  \bibinfo{author}{\bibfnamefont{N.}~\bibnamefont{Piovella}}, \bibnamefont{and}
  \bibinfo{author}{\bibfnamefont{R.}~\bibnamefont{Kaiser}},
  \bibinfo{journal}{Phys. Rev. Lett.} \textbf{\bibinfo{volume}{104}},
  \bibinfo{pages}{183602} (\bibinfo{year}{2010}),
  \urlprefix\url{http://link.aps.org/doi/10.1103/PhysRevLett.104.183602}.

\bibitem[{\citenamefont{Bachelard et~al.}(2012)\citenamefont{Bachelard,
  Courteille, Kaiser, and Piovella}}]{BachelardMie}
\bibinfo{author}{\bibfnamefont{R.}~\bibnamefont{Bachelard}},
  \bibinfo{author}{\bibfnamefont{P.~W.} \bibnamefont{Courteille}},
  \bibinfo{author}{\bibfnamefont{R.}~\bibnamefont{Kaiser}}, \bibnamefont{and}
  \bibinfo{author}{\bibfnamefont{N.}~\bibnamefont{Piovella}},
  \bibinfo{journal}{EPL (Europhysics Letters)} \textbf{\bibinfo{volume}{97}},
  \bibinfo{pages}{14004} (\bibinfo{year}{2012}),
  \urlprefix\url{http://stacks.iop.org/0295-5075/97/i=1/a=14004}.

\bibitem[{\citenamefont{Ballenegger and Weber}(1999)}]{ballenegger:599}
\bibinfo{author}{\bibfnamefont{V.~C.} \bibnamefont{Ballenegger}}
  \bibnamefont{and} \bibinfo{author}{\bibfnamefont{T.~A.} \bibnamefont{Weber}},
  \bibinfo{journal}{American Journal of Physics} \textbf{\bibinfo{volume}{67}},
  \bibinfo{pages}{599} (\bibinfo{year}{1999}),
  \urlprefix\url{http://link.aip.org/link/?AJP/67/599/1}.

\bibitem[{\citenamefont{Yang et~al.}(1999)\citenamefont{Yang, An, Perelman,
  Dasari, and Feld}}]{Yang:99}
\bibinfo{author}{\bibfnamefont{C.}~\bibnamefont{Yang}},
  \bibinfo{author}{\bibfnamefont{K.}~\bibnamefont{An}},
  \bibinfo{author}{\bibfnamefont{L.~T.} \bibnamefont{Perelman}},
  \bibinfo{author}{\bibfnamefont{R.~R.} \bibnamefont{Dasari}},
  \bibnamefont{and} \bibinfo{author}{\bibfnamefont{M.~S.} \bibnamefont{Feld}},
  \bibinfo{journal}{J. Opt. Soc. Am. A} \textbf{\bibinfo{volume}{16}},
  \bibinfo{pages}{866} (\bibinfo{year}{1999}),
  \urlprefix\url{http://josaa.osa.org/abstract.cfm?URI=josaa-16-4-866}.

\bibitem[{\citenamefont{Mackowski and Mishchenko}(2013)}]{Mishchenko}
\bibinfo{author}{\bibfnamefont{D.}~\bibnamefont{Mackowski}} \bibnamefont{and}
  \bibinfo{author}{\bibfnamefont{M.}~\bibnamefont{Mishchenko}},
  \bibinfo{journal}{Journal of Quantitative Spectroscopy and Radiative
  Transfer} \textbf{\bibinfo{volume}{123}}, \bibinfo{pages}{103 }
  (\bibinfo{year}{2013}), ISSN \bibinfo{issn}{0022-4073}, \bibinfo{note}{peter
  C. Waterman and his scientific legacy},
  \urlprefix\url{http://www.sciencedirect.com/science/article/pii/S0022407313000629}.

\end{thebibliography}

\end{document}